**Photoluminescence kinetics of dark and bright excitons in atomically thin MoS₂**

*Ilya A. Eliseyev[1], Aidar I. Galimov[1], Maxim. V. Rakhlin[1], Evgenii A. Evropeitsev[1], Alexey A. Toropov[1], Valery Yu. Davydov[1], Sebastian Thiele[2], Jörg Pezoldt[2], and Tatiana V. Shubina[1]\**

1 Ioffe Institute, 26 Politekhnicheskaya, St. Petersburg 194021, Russia
2 TU Ilmenau, Postfach 100565, Ilmenau 98693, Germany
*E-mail: shubina@beam.ioffe.ru

I. A. Eliseyev, A. I. Galimov, Dr. M. V. Rakhlin, Prof. A. A. Toropov, E. A. Evropeitsev, Dr. V. Yu. Davydov, Prof. T. V. Shubina*
Ioffe Institute, 26 Politekhnicheskaya, St. Petersburg, 194021 Russia
E-mail: shubina@beam.ioffe.ru

S. Thiele, Dr. J. Pezoldt
FG Nanotechnologie, Institut für Mikro- und Nanotechnologien MacroNano® und Institut für Mikro- und Nanoelektronik, TU Ilmenau, Postfach 100565, Ilmenau 98693, Germany

The fine structure of the exciton spectrum, containing optically allowed (bright) and forbidden (dark) exciton states, determines the radiation efficiency in nanostructures. We study time-resolved micro-photoluminescence in MoS₂ monolayers and bilayers, both unstrained and compressively strained, in a wide temperature range (10–300 K) to distinguish between exciton states optically allowed and forbidden, both in spin and momentum, as well as to estimate their characteristic decay times and contributions to the total radiation intensity. The decay times were found to either increase or decrease with increasing temperature, indicating the lowest bright or lowest dark state, respectively. Our results unambiguously show that, in an unstrained monolayer, the spin-allowed state is the lowest for a series of A excitons (1.9 eV) with the dark state being <2 meV higher, and that the splitting energy can increase several times at compression. In contrast, in the indirect exciton series in bilayers (1.5 eV), the spin-forbidden state is the lowest, being about 3 meV below the bright one. The strong effect of strain on the exciton spectrum can explain the large scatter among the published data and must be taken into account to realize the desired optical properties of 2D MoS₂.

## 1. Introduction

Atomically thin transition metal dichalcogenides (TMDs) such as MoS₂, MoSe₂, WS₂, and WSe₂ have a great potential for their use in next-generation electronic and nanophotonic devices. They exhibit strong exciton oscillator strength, an enhanced luminescence quantum yield up to room temperature, and interesting valley physics.[1,2] The key issue for realization of such benefits is the kinetics of radiative recombination,[3] which is controlled by the fine structure of the exciton spectrum comprising optically allowed (bright) and forbidden (dark) exciton states. The TMDs have a complex band structure that leads to existence of both spin- and momentum-forbidden excitons. When optically dark states have lower energies than bright ones, this arrangement reduces the radiation efficiency,[4] which is critical for many applications.

As was first demonstrated for MoS$_2$[5,6] and later adopted for other TMDs, their monolayers (MLs) are direct bandgap semiconductors, in which the lowest energy optical transitions occur at the K$_\pm$ points of the Brillouin zone. In contrast, bulk, few-layer, and bilayer (BL) TMDs are indirect gap semiconductors, in which the lowest energy transitions occur between the top of the valence band at point Γ and the bottom of the conduction band, which is either at point K or at the midpoint Λ.[7–10] Recent research has challenged this simplistic picture. In particular, momentum-indirect nature of the optical bandgap is proposed for the monolayers of MoS$_2$, WS$_2$, and WSe$_2$ based on the exciton dynamics[11,12] and resonant exciton-phonon scattering.[13] Thus, only the direct bandgap of MoSe$_2$ monolayers is beyond doubt by now.

In an atomically thin layer, the transition of band structure type from direct to indirect can naturally occur when it is subjected to tensile or compressive deformation.[14–19] The direct band gap in a MoS$_2$ monolayer is realized only in a quite narrow range of the lattice parameter deviation ($\varepsilon$) from the unstrained value ($\varepsilon$ from −1.3 to 0.3 %).[17] It should be noted that strong tensile strain arises when atomically thin TMD layers are encapsulated with hexagonal boron nitride (hBN), which is widely used to protect them from environmental influences and improve their optical properties.[20,21] In this case, direct-indirect crossover and a possible change in the energies of exciton states can complicate the interpretation of experimental data.

Strong resonances of A and B excitons, separated mainly due to spin-orbit splitting of the valence band and the influence of dielectric environment,[22] are well pronounced in the optical spectra of bulk, few-layer, and monolayer TMDs.[5,6,23–25] For a monolayer, fine structure of the exciton energy levels has been studied in detail using the methods of symmetry analysis.[26–28] In particular, the A exciton series is formed with the participation of a hole from the upper subband of the valence band and electrons from the two lowest subbands of the conduction band. It comprises the following excitons: bright (Γ$_6$), "gray" (Γ$_4$), which is an out-of-plane z-polarized state, and dark (Γ$_3$), representing an optically inactive linear combination orthogonal to Γ$_4$. The difference in energy between the Γ$_3$ and Γ$_4$ states is small, less than 1 meV;[29] therefore, with normal incidence of light on the ML, they are often collectively called "dark" (we will follow this simplification below). Several external factors may influence the fine structure of the exciton energy levels, among which strain is of particular interest.[30]

In general, the energy splitting Δ$_{AF}$ between the allowed (A) and forbidden (F) excitons can have a different sign and magnitude, since it is determined by three contributions Δ$_{AF}$ =Δ$_c$+Δ$_b$+Δ$_{ex}$, where Δ$_c$ denotes the single-particle spin splitting of the conduction band, Δ$_b$ is related to the binding energy of an exciton, mainly controlled by the spin-orbit-induced splitting, and Δ$_{ex}$ is due to the exchange interaction between the electron and hole. These contributions can have the same or opposite signs; in the latter case, they partially compensate each other. It was predicted that the lowest-energy exciton state in Mo-based materials is bright due to the opposite signs of Δ$_c$ and Δ$_{ex}$. On the contrary, the lowest state for W-based materials is dark and the splitting value can be significant (several tens of meV) due to the same sign of these contributions.[31,32]

In contrast to a monolayer, a pristine MoS$_2$ bilayer, like other TMDs with an even number of layers, is characterized by the combination of the spatial inversion and time-reversal symmetry, which lifts off spin-orbit splitting. This constraint leads to the appearance of two spin-degenerate subbands in both conduction and valence bands.[33,34] The exciton states in the entire Brillouin zone should be spin degenerate to the extent that this is determined by the spin-orbit splitting. The degeneracy of the bands can be lifted in magnetic and electric fields or under other external influences that break the symmetry.[10,33,35] In particular, helicity-resolved magneto-reflectance experiments confirmed the degeneracy and elucidated the evolution of direct bandgap K-excitons in $MoS_2$ bilayer.[36] Observation of two indirect transitions was carried out using temperature-dependent cw photoluminescence (PL),[37] but their exciton spectrum has not been clarified. The landscape of the exciton states in $MoS_2$ bilayers is enriched with interlayer excitons, which include an electron and a hole located in different layers.[36,38]

Experimental confirmation of the theoretical prediction for the excitons in monolayers was obtained by measuring the temperature dependence of the PL intensity, which shows the thermally activated contribution of the dark state to the radiation in $WSe_2$.[4,25] Polarization-resolved PL experiments with light propagating along a monolayer showed the bright-dark exciton splitting of 40 meV in $WSe_2$ and 55 meV in $WS_2$.[32] Brightening of the dark state in a magnetic field that mixes the electronic wave functions and thus makes the dark exciton states visible is a strong argument in favor of the theory's predictions. This effect was repeatedly observed for the W-based monolayers,[29,39,40] which made it possible to reliably establish their fine exciton spectrum.

In contrast to W-based materials, the experimental results on $MoS_2$ look rather contradictory. The PL temperature dependence, which is characteristic of the lower dark state, when the intensity rises with increasing temperature, was never observed in $MoS_2$ monolayers.[4,41] Moreover, the temperature dependences for the hBN-encapsulated $MoS_2$ monolayers showed that the dark trion state is 2 meV above the bright one,[42] and the neutral exciton states would likely have similar arrangement. These temperature PL dependences contradict the data of magneto-PL spectroscopy, which, in turn, are highly dispersive. Molas et al.[41] reported that the $MoS_2$ ML has a dark exciton ground state with a dark-bright exciton splitting energy of 98 meV. In a more recent work, the splitting energy in a monolayer encapsulated with hBN decreased significantly: in a transverse magnetic field, a dark exciton appeared 14 meV below the bright one under an in-plane magnetic field.[43] We emphasize that the main results and supporting information in the cited articles are very reliable; thus, additional factors must be sought that can provide such a huge spread.

Time-resolved photoluminescence (TRPL) spectroscopy could provide valuable information about exciton states in atomically thin materials. Currently, it is well established that the high oscillator strength of the direct exciton in $MoS_2$ MLs results in ultrafast radiative recombination with a characteristic time of few ps.[44,45] However, data on the radiative lifetimes of spin- and momentum-forbidden excitons in TMDs are scarce. It was predicted theoretically that the radiative decay of the dark excitons should be ~100 times slower than that of the bright ones.[46] This agrees with the measured 110-ps radiative time of the out-of-plane exciton in $WSe_2$.[29] Measurements of differential transmission transients showed that the carrier lifetime increases rapidly from ~50 ps in a ML to 1 ns in 10 layers of $MoS_2$,[47] which may reflect the contribution of momentum-forbidden transitions in multilayer indirect-gap structures. Even in a monolayer, the radiation window inside the light cone is very narrow in TMDs,[48] and excitons leaving the window must also have a long decay time, since they are indirect in momentum. It should be noted that the time domains, in which previous TRPL measurements were carried out, might be not wide enough to correctly determine the long decay times characteristic of various dark states. Thus, the TRPL method must be adapted to analyze the rich landscape of excitons in TMDs.

In this paper, we present experimental evidences that the spin-allowed bright exciton is the lowest in the series of A-excitons in $MoS_2$ monolayers and bilayers that are unstrained or under compressive strain; on the contrary, the spin-forbidden exciton is the lowest in the momentum-indirect exciton series (1.5 eV) in bilayers. This conclusion is based on detailed time-resolved micro-photoluminescence measurements in a wide temperature range (10–300 K) and in an extended time domain (25 ns) of samples on planar and patterned substrates, in which the strain values were determined using a micro-Raman study. The temperature dependencies of characteristic PL decay times, which we assigned to the spin-allowed, spin-forbidden, and momentum-forbidden excitons, are non-monotonic; they begin to either increase or decrease at certain temperature due to a change in the thermalized population of the upper state. This behavior makes it possible to estimate the sign of the bright–dark exciton splitting (negative when the dark is upper) and its magnitude. These estimates are additionally confirmed by the temperature-dependent contribution of the corresponding component to the integrated emission intensity. The splitting energy of A excitons in unstrained samples turns out to be rather low (< 2 meV); however, in structures subjected to compressive deformation it increases several times. Our data unambiguously show a strong effect of strain not only on the band structure as a whole, but also on the fine spectrum of exciton states, which can explain the scatter of literature data on this matter.

## 2. Samples

Atomically thin $MoS_2$ flakes were mechanically exfoliated from a bulk crystal and transferred either onto planar substrates, $SiO_2/Si$ and $Si_3N_4/Si$, or onto a patterned $Al_2O_3$ substrate with pyramids on their surface, as described in the section 7.1 of Methods. **Figure 1** shows optical images of the structures fabricated on the patterned (a) and planar (b) substrates. The regions of monolayer and bilayer thicknesses were determined by micro-Raman spectroscopy, described in Methods, 7.2. Focusing the laser beam into a spot with diameter of <1 μm allows us to measure separately the regions of different thickness and to focus on the top of the pyramids and in interspaces between them.

A non-resonant Raman spectrum of high-quality $MoS_2$ in the range of 50-500 cm⁻¹ consists of two main lines usually referred to as $E_{2g}$ (380 cm⁻¹) and $A_{1g}$ (405 cm⁻¹), and an asymmetric $2LA$ peak (Figure 1, (c)). The $E_{2g}$ and $A_{1g}$ lines correspond to the in-plane and out-of-plane vibrations of the S atoms, respectively,[49] while the $2LA$ line is an overtone of the longitudinal acoustic mode.[50] Difference between the $E_{2g}$ and $A_{1g}$ lines positions is a well-known way of determining the number of layers. The typical $\omega(A_{1g})$-$\omega(E_{2g})$ frequency difference lies in the range of 18-20 cm⁻¹ for monolayer[51–54] and 21-22.3 cm⁻¹ for bilayer[51,54] $MoS_2$. However, positions of these lines can be affected by different factors such as strain and doping,[51,54] thus, the most unambiguous way to distinguish between the monolayer and bilayer is to analyze the low-frequency part of the spectrum (0–50 cm⁻¹). In case of bilayer, the in-plane and out-of-plane motion of the $MoS_2$ layers with respect to each other cause the appearance of shear ($C$, 23 cm⁻¹) and layer-breathing ($LB$, 40 cm⁻¹) modes, respectively.[55] In the spectrum of monolayer $MoS_2$ these lines are obviously absent due to the lack of the second layer.

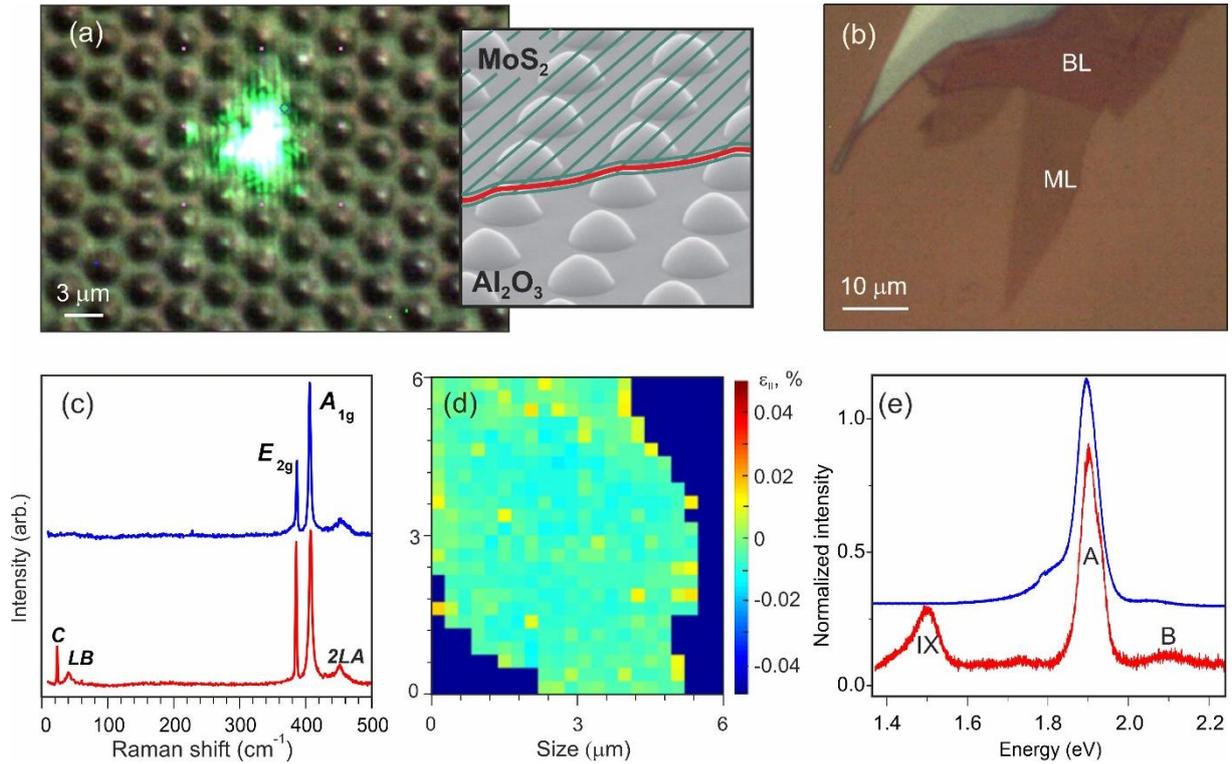

**Figure 1.** (a, b) Optical microscope images of atomically-thin MoS₂ samples transferred onto different substrates: (a) patterned sapphire and (b) planar SiO₂/Si with indication of monolayer (ML) and bilayer (BL) regions. The bright green spot in (a) is a focused laser beam; the inset shows a bird-view scanning electron microscopy image of the patterned substrate with a schematic MoS₂ film. (c) Raman spectra measured in ML (blue line) and BL (red line) regions. (d) Distribution of strain along a planar MoS₂ monolayer on a SiO₂/Si substrate, confirming that it is unstrained. (e) Typical PL spectra of a ML (blue line) and a BL (red line) with peaks of direct A and B excitons and indirect IX exciton (77 K). The Raman and PL spectra are normalized to the maximum intensity and vertically shifted with respect to each other for better visibility.

Figure 1 (c) shows typical Raman spectra measured in the different areas on the planar structures on the Si-based substrates. In the areas denoted as bilayer, the $C$ and $LB$ lines are present on their classical positions, and the frequency difference ranges from 21.7 cm⁻¹ for MoS₂ on Si₃N₄/Si to 22 cm⁻¹ on SiO₂/Si. This allows us to unequivocally determine the bilayer areas. In other spectra, the $C$ and $LB$ lines are absent in the low-frequency part, and the frequency difference $\omega(A_{1g})$-$\omega(E_{2g})$ is 19.8 cm⁻¹ for MoS₂ on the planar substrate and 18.3 cm⁻¹ on the patterned substrate. This verifies that MoS₂ is a monolayer in corresponding areas. Analysis of the Raman mapping data (details of the analysis are described in the next section) obtained on MoS₂ monolayers on the planar substrate shows that the areas under study are practically strain free (Figure 1 (d)).

Micro-photoluminescence spectra are measured in monolayer and bilayer regions using cw excitation by a 532-nm line of a Nd:YAG laser (Figure 1 (e)). They are fully consistent with the published spectra of MoS₂ samples placed on different substrates.[5,6,56] The peak of the A exciton dominates the spectra in the monolayers, while the peak of the B exciton is markedly smaller. An additional peak associated with the indirect exciton IX appears at 1.5 eV in bilayers. In the thicker parts of the flakes (3-layers and more), this IX peak is shifted towards 1.3 eV. Thus, such measurement of PL spectra is an independent way to confirm our determination of the atomically-thin regions.

## 3. Strain in atomically-thin MoS$_2$

We perform the combined micro-Raman and micro-PL measurements to analyze the strain in the structures under study. For our planar samples, the $E_{2g}$ and $A_{1g}$ modes are found at 384.3 and 404.4 cm$^{-1}$ in MLs, while in BLs they are at 383.6 and 405.3 cm$^{-1}$. In the limits of experimental accuracy, these values are in good agreement with previously published data for structures with negligible strain and doping levels.[16,49,53,57] Therefore, we assume our planar samples as strain- and doping-free and take these mode frequencies as references to determine the strain in ML and BL on the patterned substrate. The correlation analysis of the $A_{1g}$ and $E_{2g}$ line positions, that allows one to separate the strain and doping contributions to the shift of these lines, was carried out using the expressions (1) and (2) from the work of Kim et al.[54] We used the Gruneisen parameters for the case of biaxial strain and electron density shift rates for monolayer and bilayer MoS$_2$ given in this paper. By this analysis, we derived the values of carrier concentration and biaxial strain for each spectrum. For planar and patterned monolayers and bilayers, the estimated doping level varied in the range $2\text{-}4\cdot10^{12}$ cm$^{-2}$, which is low in comparison with the values that can provide a noticeable shift of the PL lines.[58] According to the Raman mapping analysis, the scatter of the strain values for planar monolayer in the area of interest was negligible, $\varepsilon_\parallel=(0\pm0.01)\%$, (see Figure 1 (d)), and the same statement holds for bilayer (map not shown here).

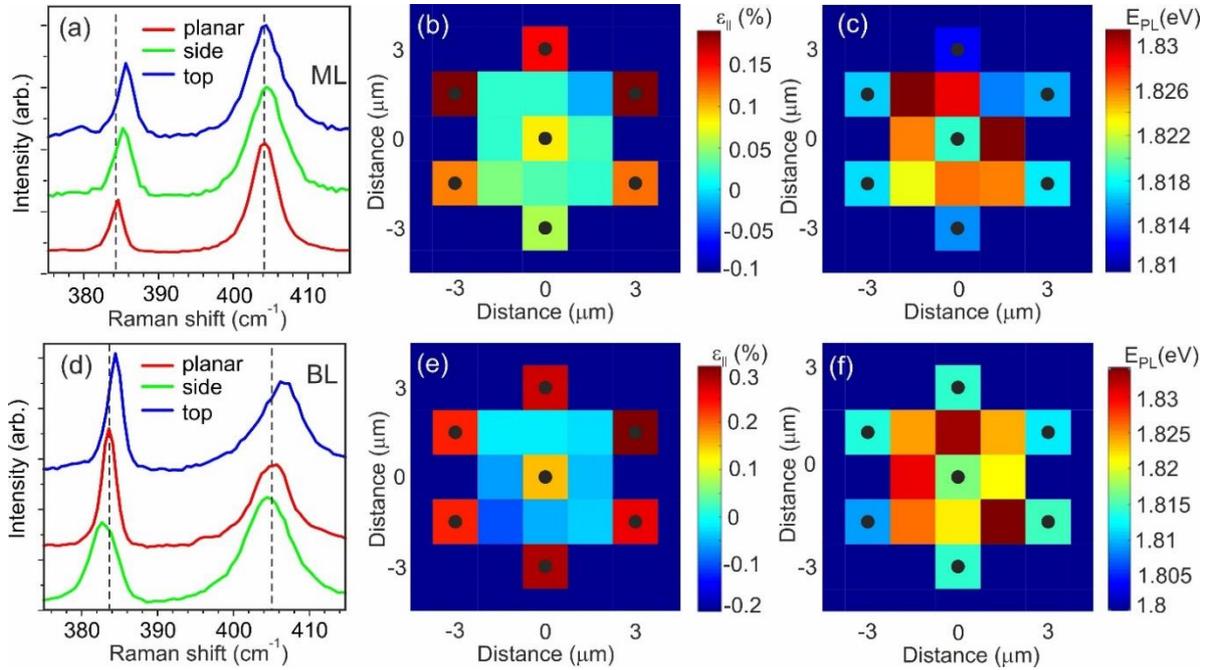

**Figure 2.** (a, d) Typical Raman spectra of (a) ML and (b) BL on a planar substrate and on a patterned Al$_2$O$_3$ substrate recorded focusing on pyramids (top) and between them (side). (b, e) Mapping of biaxial strain values $\varepsilon_\parallel$ derived from the Raman mode energies for (b) a monolayer and (e) a bilayer on a patterned substrate. Black spots mark the top of pyramids. The negative $\varepsilon_\parallel$ sign corresponds to the tensile strain, the positive – to the compressive one. (c, f) Mapping of the PL energy distribution along the (c) monolayer and (f) bilayer, measured in the same spots as the Raman spectra. All measurements are carried out at 300 K.

The site-selective micro-Raman spectra of ML and BL measured in the pyramids and between them are shown in **Figure 2** together with those of the reference planar samples. Previous experimental studies of the samples, strained mechanically,[16,52,53] showed that the tensile strain provides a lower-energy shift of both $E_{2g}$ and $A_{1g}$ modes. For the more strain-sensitive $E_{2g}$ mode, the shift induced by the tensile deformation is about $-5$ cm$^{-1}$/%. Under compressive strain, the modes are shifted to the higher energy.[59] The Raman data obtained in our experiments indicate that at the tops of the pyramids the MoS$_2$ monolayer is compressed by an average of 0.12% compared to the planar one, while the compressive strain in the bilayer is higher, about 0.28%. It is useful to compare these data with those obtained by similar Raman studies of the MoS$_2$ ML and BL, where the strain appears after hBN encapsulation. The derived strains are -0.06 and -0.29%, respectively,[20] being of the same order of magnitude, but with a negative sign due to tensile deformation. Thus, we can conclude that the strain is not caused by the lattice mismatch with a substrate, as in conventional semiconductor heterostructures, but rather by fabrication methods in both cases. A possible reason of the greater strain in the bilayers is that the thicker a film, the more difficult it is for it to relax elastically.

With selective study of μ-PL in the tops and interspaces between the pyramids, we have faced an intricate situation. Everywhere in the compressively strained regions, the energy of PL peak was shifted to the lower energy with respect to the planar strain-free structure. At room temperature with 532-nm line excitation, it was on average at 1.815 eV on the top of the pyramids, while on the flat substrate it was at 1.837 eV. There is also a difference between the regions suspended and contacted with the Al$_2$O$_3$ tops (Figure 2 (c, f)), which is higher than can be induced by a change in the dielectric environment.[60] Note that these areas belong to the same flake, where the carrier concentration is uniform.

Generally speaking, a shift of the PL energy can occur due to modification of the band structure, which can be caused by both tensile and compressive strain. In a monolayer under tensile strain, the valence band at the Γ point rises up, and the conduction band in the K point shifts downward. As a result, the momentum-indirect transition KΓ turns out to be the lowest in energy. Under compression, the minimum of the conduction band is in the Λ point, while the valence band at the Γ point goes down, and the lowest-energy transition is ΛK. Thus, the MoS$_2$ monolayer is undoubtedly direct-gap semiconductor only within a rather narrow strain range, in which the PL line exhibits a linear red shift under tension and weaker blue shift under compression (Figure 3 (b)). Outside this range, compressive deformation can lead to a decrease in the PL energy due to the admixture of the indirect ΛK transitions.

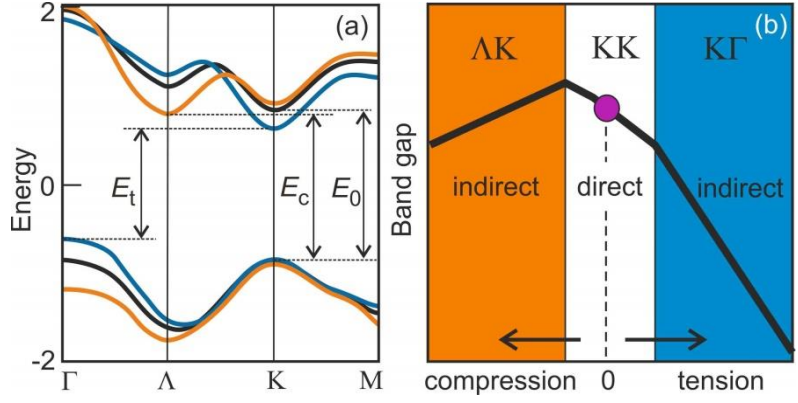

**Figure 3.** (a) Sketch of the band structure in a MoS$_2$ monolayer: without strain (black), under compressive (orange) and tensile (blue) strains less than 2% (based on Refs.[12,13,15]). $E_0$, $E_c$, and $E_t$ denote the corresponding band gap energies. (b) Modification of the band gap from direct to indirect for various types of deformation.

Theoretical studies predicted that such a decrease in the optical gap energy will take place at a compression deformation of about 1-2%,[14–17,61] although the calculation of the strain-induced tunability demonstrated the direct-to-indirect band gap crossover at ~0.3%,[62] which is closer to our case. It should be noted that the behavior of emission in an atomically thin structure under compression is much less studied than for structures under tension, for which there are many publications.[16,53,63–65] This is because the compressive strain is more difficult to achieve experimentally. As far as we know, the only experiment was the study of trilayer MoS$_2$, located on a piezoelectric substrate, providing a compression of ~ 0.2%.[66] Neither monolayers nor bilayers were investigated experimentally under compression.

Among other factors influencing the PL spectra, there is the use of excitation above the band gap, which increases the population of carriers in the upper valleys.[67] This leads to a quasi-equilibrium carrier distribution instead of a non-equilibrium one in the K valleys with the quasi-resonant 532-nm excitation. In a MoS$_2$ monolayer, such an excitation caused the blue-shift of the PL peak by 30 meV relative to its position in the case of the quasi-resonant excitation by a 532-nm laser line and a suppression of the B exciton line.[67] Similar effects are observed in our structures upon the 405-nm excitation. In addition, Steinhoff et al.[67] demonstrated that the 532-nm excitation is more effective, because it provides a larger number of photo-induced carriers at the same excitation power. In general, it can lead to a red shift of the emission. The combined action of these effects can explain the ~40 meV energy difference in our spectra measured in the same structure at different excitation wavelengths. As also shown in Ref.,[67] at the carrier concentration of >10$^{13}$ cm$^{-2}$, the influence of strain on the band structure is more pronounced. However, this concentration is markedly higher than in our samples, and we can consider the strain effect alone. In particular, we expect that the strain can change the fine spectrum of excitons, namely: the energy splitting between the bright and dark states, similar to strain-induced tuning of the quantum levels in the quantum dots.[68]

## 4. Time-resolved micro-photoluminescence studies

The primary task of our research was to determine the splitting of dark and bright excitons in unstrained films. The width of the low-temperature line in our monolayer (~30 meV) does not allow distinguishing the contributions of the dark and bright states, since the separation between them can be small.[31] However, for temperature-dependent TRPL spectroscopy, it does not matter how close they are, since this method deals with the characteristic times of rapidly and slowly decaying components presumably associated with these states. Monitoring the changes in decay times with temperature makes it possible to clearly determine the sign of the splitting and approximately estimate its magnitude.

The time resolved micro-PL studies of the $MoS_2$ samples on flat and patterned substrates were carried out using an ST-500-Attocube cryostat (Janis) supplied with a three-coordinate piezo-driver, which allows maintaining the position of the laser spot on a sample during a long-term series of the measurements with an increase in temperature from 10 to 300 K. Focusing through the cryostat window was done using a 50x Mitutoyo objective (NA = 0.42). Therefore, the signal was collected from a spot with a diameter of ~4 μm that exceeds the distance between the pyramids. The 405-nm line of a pulsed laser was used for excitation to probe reliably the whole set of exciton transitions. In our TRPL experiments, the chosen average power of excitation was as low as ~0.1 μW and a laser spot had a diameter of 3-5 microns; the minimal size corresponds to the average power density of ~0.01 μW·μm$^{-2}$. This excludes the changes in spectra under laser exposure, which were observed at the power of 2-3 orders of magnitude higher.[69] Under such conditions, the estimated concentration of photo-induced carriers in a monolayer is ≤10$^{10}$ cm$^{-2}$ that is markedly less than the intrinsic carrier concertation. The enhancement of trion formation due to photodoping is negligible. In our measurements, the spectral width of the filter window (10 nm) was chosen to record the predominantly exciton part. (For more details, see Methods).

Let us first consider the monolayers placed on the different substrates. The temperature evolution of the spectra in the vicinity of the A exciton in the planar monolayer is shown in **Figure 4 (a)**. We observe the red-shift of ~50 meV and broadening of the A exciton peak, similar to what was reported for other 2D TMDs.[25,70,71]

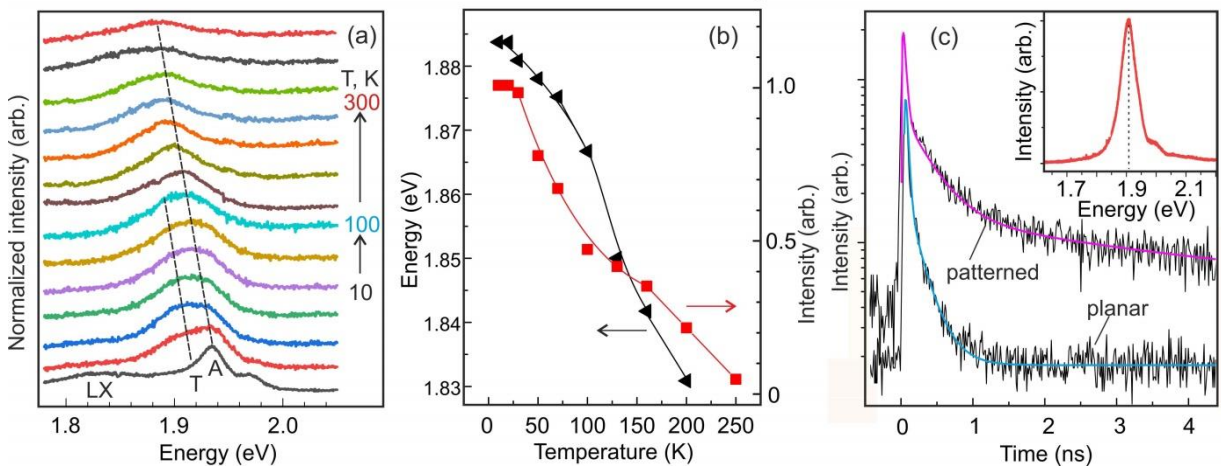

**Figure 4.** (a) Temperature variation of the normalized PL spectra (vertically shifted for clarity) measured in a $MoS_2$ monolayer on a planar $SiO_2/Si$ substrate using 405-nm pulsed excitation with a power density of 6 W/cm$^2$. A and T denote the A exciton and trion, LX – localized exciton. (b) Temperature dependencies of the energy and intensity of the PL peak recorded in a monolayer on a patterned $Al_2O_3$ substrate with cw excitation by a 532-nm line.

(c) The onset part of PL decay curves (black lines) in the monolayers placed on different substrates measured at 10 K at 6 W/cm²; the color lines present a fitting with the decay times as follows: $MoS_2/SiO_2 - t_2 = 0.22$ ns, $t_3 = 0$; $MoS_2/Al_2O_3 - t_2 = 0.22$ ns, $t_3 = 3.1$ ns. The inset shows a spectrum recorded at 10 K in the patterned sample during this TRPL measurement.

At low temperatures, this peak has shoulders at both sides. The nature of higher-energy shoulder, probably related to excited states of localized excitons whose band is situated at ~1.8 eV, is still under investigation. The trion contribution is almost absent at 10-20 K, then it starts to increase, resulting in the broadening of the PL line. We assume that this effect can be associated with the temperature-induced escape of the carriers from weakly localizing sites. The disappearance of the broad peak of localized excitons, LX, is consistent with this assumption. The enhanced carrier concentration promotes the formation of trions above 20 K. Eventually, the neutral exciton line survives up to room temperature due to giant oscillator strength, while the trion dissociates to ~100 K, as in other TMD structures with the modest carrier densities of ~$10^{12}$ cm$^{-2}$.[72-74]

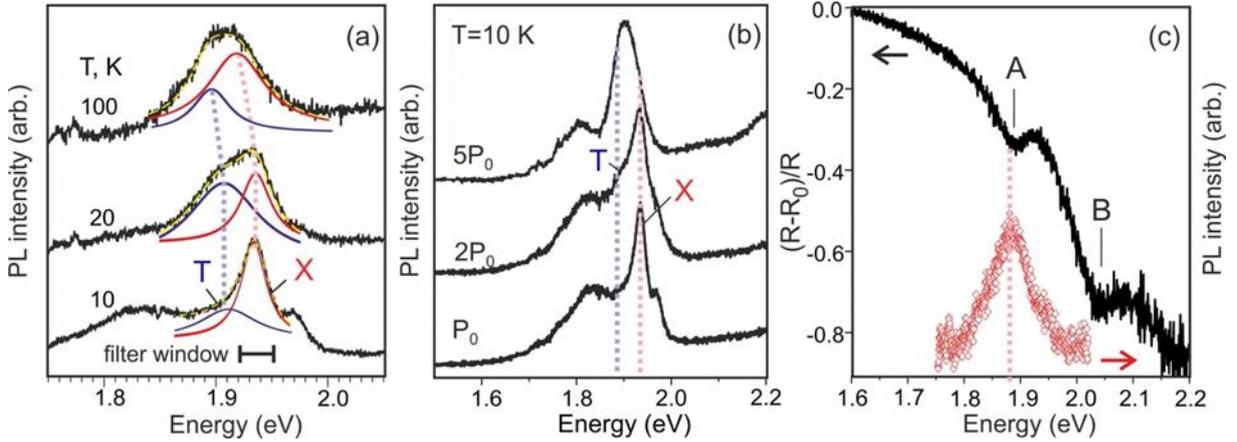

**Figure 5**. (a) PL spectra of the $MoS_2/SiO_2$ monolayer, measured at the same place at different temperatures during the TRPL studies, shown together with their deconvolution over two Lorentzians. (b) Low-temperature PL spectra measured at this place at different excitation powers ($P_0 = 100$ nW, excitation 405 nm, filter window 10 nm). (c) Spectra of differential reflectivity and PL measured at room temperature in the same $MoS_2$ monolayer.

To confirm the correct assignment of the emission features, we show in **Figure 5** (a) the deconvolution of PL spectra measured at different temperatures. The separation between the exciton and trion components perfectly corresponds to the commonly accepted trion binding energy of ~30 meV in $MoS_2$.[42] The variation of spectra with increasing power is shown in Figure 5 (b). The intensity of the exciton emission rises linearly, while for trion it increases super-linearly.[75] The PL broadens and shifts towards the position of trion when the photo-induced carrier density is increased with increasing power. In Figure 5 (c), the A exciton resonance in a reflectivity spectrum perfectly correlates with the main peak in the PL spectrum. The resonance of trions is not pronounced due to their full dissociation at room temperature. We reject the localized exciton emission as the origin of the dominating peak, because the localized exciton should inevitably quench with increasing temperature via interaction with phonons.[70,76]

In the monolayer on the patterned substrate, the A exciton peak has an average width of 60 meV at 10 K. The quenching of its intensity is not monotonic (Figure 4 (b)). The inset in Figure 4 (c) shows that it is redshifted in energy compared to the planar monolayer PL. In addition, we observe the suppression of the B exciton line with the non-resonant 405 nm excitation, predicted in Ref.[67] Strong difference is observed in the decay curves measured in the A peak of these two samples: the slowly decaying component appears in the monolayer on the patterned substrate, while it is absent in the planar monolayer.

The time resolution of used system, $t_0$, was about 50 ps; thus, we were unable to resolve the very short radiative time of the A exciton, which is only few ps.[3,44,45] However, we can track the peak amplitude immediately after excitation, which gives us information on how the fast component changes with temperature. To take into account the instrumental function, we simulate this sharp peak by two rising (r) and decaying (d) exponents with the same amplitude $A_1$ and characteristic times $t_1^{r,d} \sim t_0/2$, as $A_1 \cdot (\exp(-t/t_1^d) - \exp(-t/t_1^r))$, where $t_1^d$ was slightly longer that $t_1^r$. The integral intensity of this rapidly decaying contribution is assumed to be $I_1 \approx A_1 \cdot t_1$. To model the rest, we used two exponents with middle, $t_2$, and slow, $t_3$, decay times. We also accounted a background contribution, which was extremely slow at every temperature, by subtracting that as a constant. Thus, the change in the total intensity with temperature is given by the expression $Sum = A_1 \cdot t_1 + A_2 \cdot t_2 + A_3 \cdot t_3$, and the contribution of a particular component is equal to $A_i \cdot t_i / Sum$.

Modeling the decay curves measured at different temperatures in monolayers on planar and patterned substrates gives the results shown in **Figure 6**. In the unstrained monolayer on $SiO_2/Si$, the slow component characterized by time $t_3$ is absent. In contrast, it exists in the strained monolayer on the $Al_2O_3$ substrate. For us, this is a weighty argument in favor of assigning the time $t_3$ to the momentum-indirect transitions which can be realized in strained structures. Components with the intermediate decay time $t_2$ exist in both samples, although demonstrate different variation with temperature. In the previous studies of 2D TMDs, the lifetime of 0.1-0.3 ns was associated with either neutral or charged dark excitons.[29,77] Since the emission of trions is cut by the interference filters, we can regard $t_2$ as the characteristic time of spin-forbidden dark exciton states, observed up to room temperature.

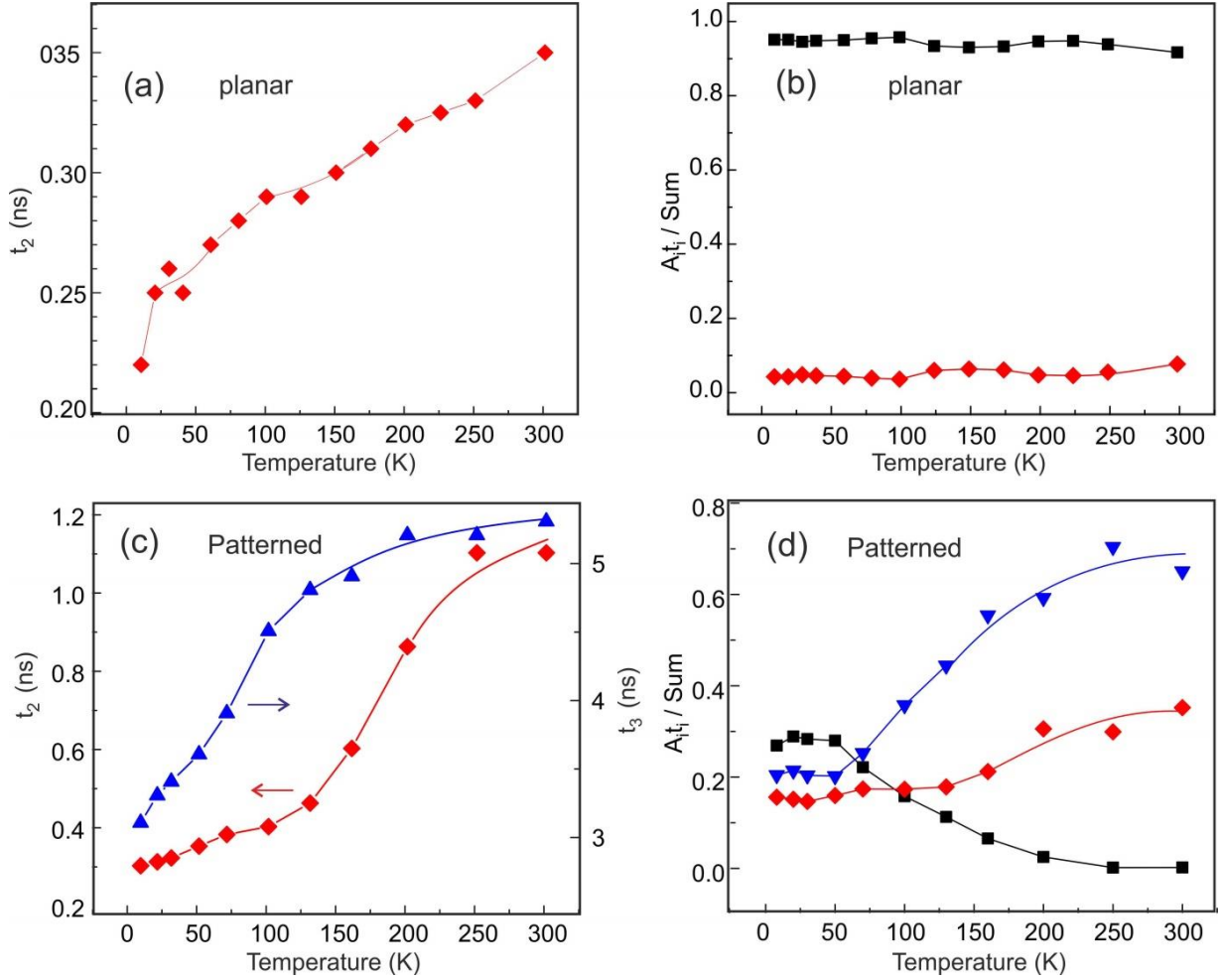

**Figure 6.** (a,c) Characteristic decay times $t_2$ and $t_3$ for the monolayers on the (a) SiO$_2$/Si planar and (c) Al$_2$O$_3$ patterned substrates. (b, d) Relative contributions of the fast $A_1 \cdot t_1$ (black squares), middle $A_2 \cdot t_2$ (red diamonds), and slow $A_3 \cdot t_3$ (blue triangles) components to the total radiation for the monolayers on the planar (b) and patterned (d) substrates. The lines are guides for the eyes.

In both MoS$_2$ monolayers under study, the characteristic times $t_2$ and $t_3$ increase with a temperature rise. At low temperatures (<100 K), it can be interpreted in terms of an increase in the population of dark states as observed in various nanostructures.[48,78–81] In the first approximation, the longer characteristic time $\tau_L$ in each doublet, comprising optically allowed and forbidden by spin or momentum states, can be described via their recombination rate at low temperature:[79]

$$\tau_L^{-1} = \frac{\Gamma_A + \Gamma_F}{2} - \left[\frac{\Gamma_A - \Gamma_F}{2}\right] \cdot \tanh\{\Delta E/(2k_B T)\}, \quad (1)$$

where $k_B$ is the Boltzmann constant, $T$ – temperature, $\Gamma_A$ and $\Gamma_F$ are the respective recombination rates of the allowed and forbidden states, and $\Delta E$ is the energy of acoustic phonons which matches the dark-bright exciton splitting $\Delta_{AF}$ (here, it is assumed that thermalization of exciton states is via the interaction with the phonons). When $k_B T > \Delta E$, the second term in Equation 1 decreases, and the recombination rate tends to the half-sum of $\Gamma_A$ and $\Gamma_F$ at low temperature. Accordingly, $\tau_L$ will increase when the bright exciton is the lowest state possessing a high recombination rate; and decrease when the lowest state is dark. The temperature value corresponding to a threshold change in the populations of states can be used to estimate the splitting of a bright and dark excitons, assuming $\Delta_{AF}=E_A-E_F\approx k_BT$, while its sign will be determined either by an increase ("-") or decrease ("+") of $\tau_L$.

In the unstrained structure, the sharp increase in the time $t_2$ at 10-20 K (Figure 6 (a)) corresponds to small value $\Delta_{AF}\approx$-(1-2) meV with the bright exciton being the lowest. There is some uncertainty in determining the absolute value of this splitting due to the relatively small distance between the dark and bright states, but the slope of the temperature dependence of $t_2$ undoubtedly indicates their mutual arrangement. Additional argument for confirmation of the above assumption is the dominant contribution of the fast component to the total intensity over the entire temperature range (Figure 6 (b)). Similar picture is also observed for the direct A exciton in a bilayer on a planar substrate (not shown here).

The strain in the monolayer on the patterned substrate causes pronounced changes (Figure 6 (c), (d)). The temperature dependence of the $t_2$ value has two pronounced kinks near 30-50 K and around 100 K. Since the contribution $A_1 \cdot t_1$ dominates not up to 100–130 K, but only up to 50 K, when the $A_2 \cdot t_2$ component begins to rise, we accept $\Delta_{AF} \approx$-4 meV. Although the relative contribution of the $A_1 \cdot t_1$ component to the sum intensity rapidly decreases due to the increase of $A_2 \cdot t_2$ and $A_3 \cdot t_3$ components, its absolute value is stable up to ~100 K, which indicates the neglecting nonradiative process in the temperature range of 10-100 K. The observed increase in $t_3$ and $A_3 \cdot t_3$ can be associated with a direct-indirect crossover of the band gaps in monolayers, similar to that observed in a multilayer structure.[82] It means that the energy difference between the transitions K-K and K-$\Lambda$ is decreased under compressive strain. At higher temperatures, the momentum-indirect excitons which are out of the light cone can contribute to the slow radiation.

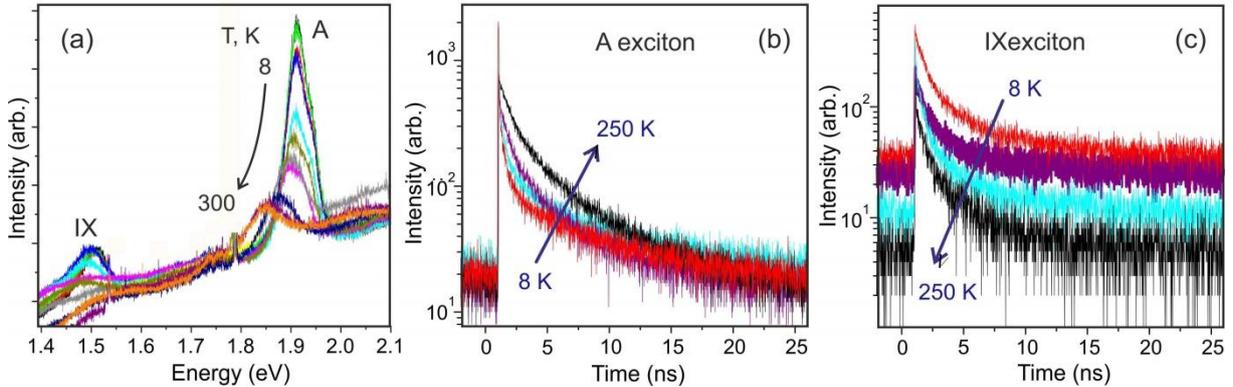

**Figure 7.** (a) Temperature variation of the PL spectra measured in a MoS$_2$ bilayer on a patterned Al$_2$O$_3$ substrate using 405-nm excitation (the slope is due to the contribution of higher exciton states with above-barrier excitation). (b, c) Selected decay curves measured in the bilayer for the (b) A exciton (1.9 eV) and (c) indirect IX exciton (1.5 eV).

A completely different picture is observed in radiation kinetics of a highly stressed bilayer on a patterned substrate (**Figure 7**). This more complex system initially includes both direct and indirect, IX, excitons. With increasing temperature, the slow component in the PL decay curves increases for the A exciton and disappears for IX. The modelling shows that, in contrast to the monolayer on the same patterned substrate, variation of $t_2$ and $t_3$ for the A exciton transition is opposite with the temperature rise: $t_2$ increases, while $t_3$ decreases. The onset of the $t_2$ rise occurs at ~100 K that corresponds to $\Delta_{AF} \approx -8$ meV with the lowest bright exciton. At the same time, the $t_3$ value gradually decreases from 6.3 ns to 4.5 ns starting from 30-40 K. For the indirect IX transition, both $t_2$ and $t_3$ times are decreased. It should be noted that for the reference bilayer on the planar substrate (not shown here), no variation in $t_2$ times with temperature was observed ($\Delta_{AF} = 0$), which is in good agreement with the predicted degeneracy of conduction band due to even number of monolayers.

As the temperature rises, the contribution of the fast $A_1 \cdot t_1$ component of the A exciton dominates the total radiation intensity up to 175 K, when the combined action of the slower components begins to prevail and the nonradiative recombination is activated. On the contrary, for an indirect IX exciton (**Figure 8**), a noticeable contribution of the $A_1 \cdot t_1$ component appears at high temperatures, when an approximately twofold decrease in both $t_2$ and $t_3$ leads to a decrease in their contributions to the total intensity. In a planar bilayer, the indirect IX transition also shows a decrease in both $t_2$ and $t_3$, starting at 30 K. The much weaker component $A_1 \cdot t_1$ slightly increases at this temperature ($\Delta_{AF} < 2$ eV). Based on these data, the spin-forbidden state is assigned as the lowest in the indirect exciton series (1.5 eV) in both strained and unstrained bilayers.

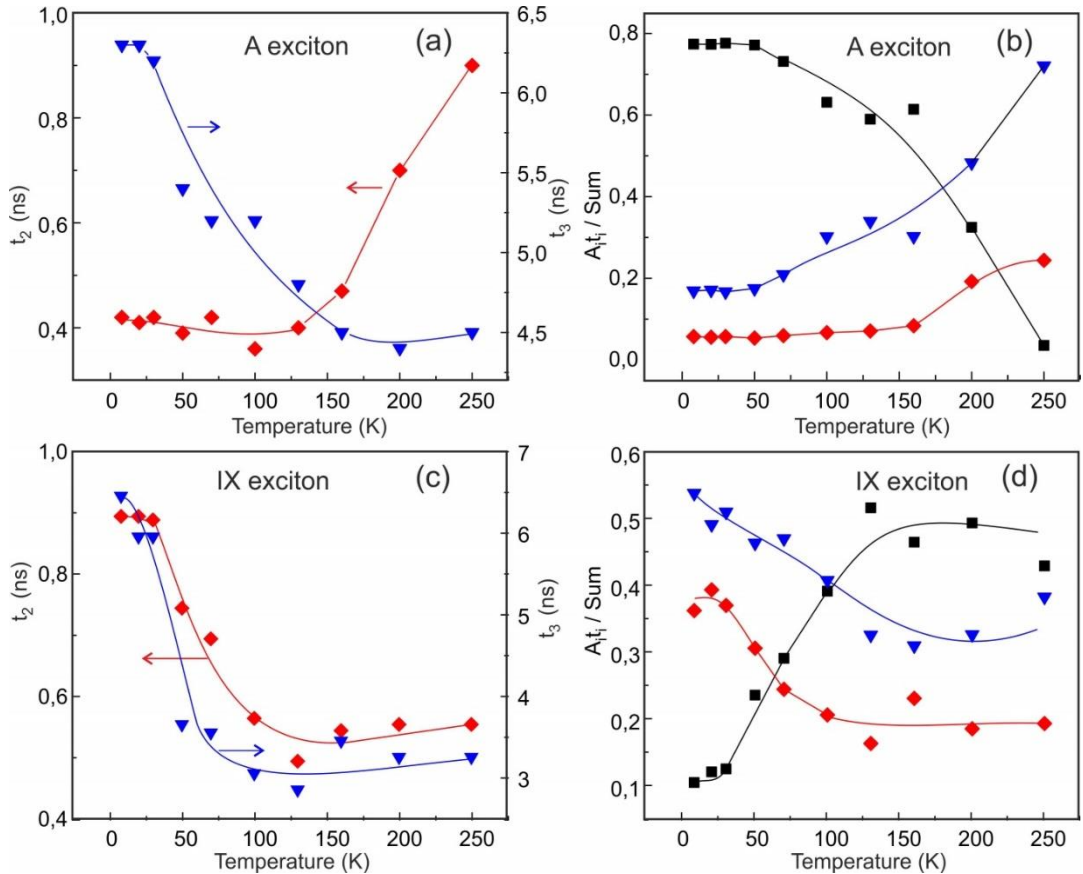

**Figure 8**. (a,c) Characteristic decay times $t_2$ and $t_3$ derived for (a) the A exciton and (c) the indirect IX exciton transitions in the bilayer on the patterned substrate; (b, d) represent the relative contributions of the fast $A_1 \cdot t_1$ (black squares), middle $A_2 \cdot t_2$ (red diamonds), and slow $A_3 \cdot t_3$ (blue triangles) components to the total radiation for these transitions in the bilayer. The lines are guides for the eyes.

## 5. Discussion

In this work, we analyze the mutual arrangement of the dark and bright excitons in the MoS₂ monolayer and bilayer, which are either not strained or are subjected to compressive strain. The schematic of the possible states is given in **Figure 9,** which includes the spin- and momentum forbidden excitons of different valleys. According to DFT calculations,[10] there are two exciton series, K−K and Λ−K, for the monolayer and three, K−K, Λ−Γ, and Λ−K, for bilayer. Under strain, one should expect the changes in energies of these series and in the exciton splitting inside them.

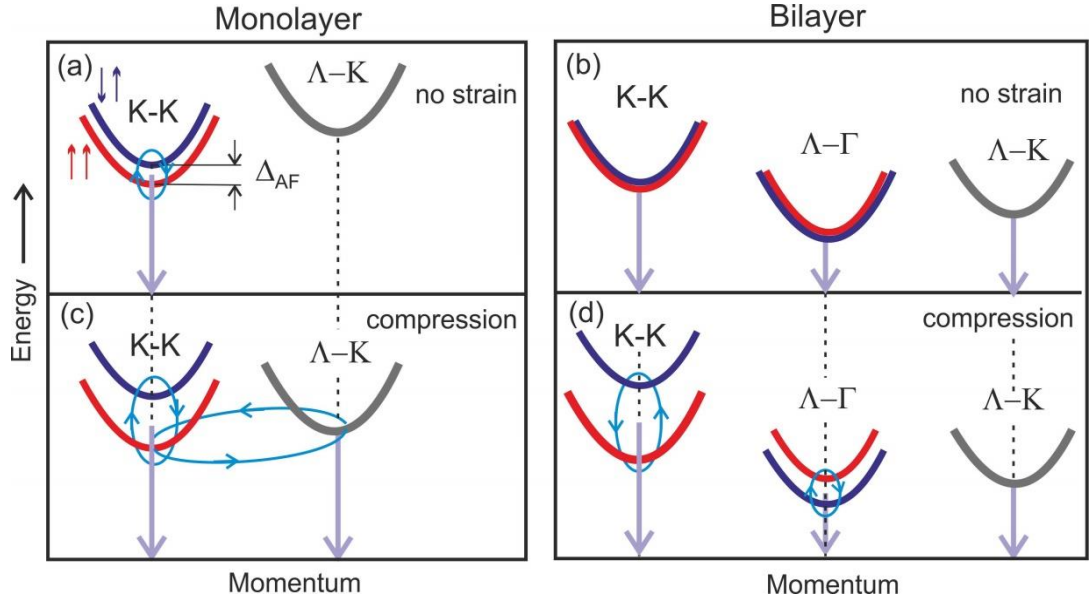

**Figure 9**. Exciton states in unstrained (a, b) and compressively strained (c, d) monolayers and bilayers. Without strain, the ordering of exciton states is shown in accordance with Ref.[10] The K-K exciton states are shown for the K+ (allowed states spin-up). The bright and dark states are shown by the red and blue lines; gray lines mark Λ−K states with unknown excitons. Vertical arrows indicate radiative transitions.

We consider the results of the temperature-dependent TRPL using a simplified model for a system containing dark and bright states, assuming that when the dark exciton is at the top, the values of decay time constants increase with increasing temperature; if it is the lowest, the time is shortened.[79,83] The temperature when the states begin to thermalize allows to estimate approximately the dark-bright energy splitting. In addition, we take into account the weights of components with different decay times in the total PL intensity. When the fast component dominates over the whole temperature range, this is a signature of the ground bright state. In the case of MoS₂ monolayers, this combined approach gives a distinct result: the bright exciton is the lowest and the bright-dark splitting energy does not exceed 2 meV when the strain is absent. This result is in accordance with many theoretical predictions (see references in Ref.[27]) and previously reported PL intensity dependencies on temperature.[4,41] The dependences for the K-K exciton in a planar bilayer, in which no splitting was found, fully agree with the predicted degeneracy of the conduction and valence bands for the structure with even number of monolayers.[10,33,34] These findings confirm the validity of our approach.

The feature of 2D TMDs is the very high recombination rate of the bright K-K exciton, which can efficiently recombine even when the dark state is the lowest.[29,43] This makes questionable achieving the full thermal equilibrium in the exciton series. Conditions for that in TMD monolayers were analyzed by Zhang et al.[4] They showed that the characteristic time of exciton thermalization is within the time scale of tens of picoseconds that is relevant for our measurements. The rapid radiative decay might result in certain depletion of the exciton population within the radiative cone. However, these exciton states constitute only a relatively small fraction of all excitons for the temperature range of interest and the overall exciton energy distribution should remain largely thermal despite the rapid decay within the radiative cone.

Temperature dependent nonradiative recombination can also affect the exciton dynamics. To evaluate its impact, we examined the temperature variation of the fast component $A_1 t_1$ of the K-K exciton. In the temperature range of interest to us (10-100 K), where bending and an increase in temperature dependences are observed, the intensity of this component is practically constant in the planar monolayer and in the bilayer. This means that the internal quantum efficiency in this temperature range is high, and nonradiative channels do not significantly affect our results. In patterned samples, after 100 K, the $A_1 t_1$ intensity decreases stronger than in the planar samples. Possibly, providing compressive strain in structures above an $Al_2O_3$ pyramid leads to the generation of some additional defects that serve as centers of nonradiative recombination.

Another important factor is the influence of excitons outside the light cone, which need the help of phonons to recombine. It was previously shown that the presence of such states in TMDs can increase the emission time from several ps to 1 ns.[4] However, most of the integral PL intensity in the planar structures is due to the fastest component, the decay time constant of which does not change within the time resolution of ~50 ps in our setup over the entire temperature range. A similar behavior was previously observed in $MoS_2$ monolayers by Lagarde et al.[44] It can be assumed that the contribution of the out-of-cone excitons was assigned by us to the slowest momentum-indirect component, which increases with temperature. However, it cannot explain the fast times of the dominant component of the emission, which persist up to high temperatures. Additional research is needed to elucidate this inconsistency.

To find out how strain affects the spectrum of the exciton states, we realize compression in the layers under study in an original way — by placing a film on a patterned substrate with pyramids. The results of TRPL studies of such structures clearly demonstrate that the state ordering in the A exciton series is the same (the bright is the lowest) but the $\Delta_{AF}$ value can increase several times. Importantly, the arrangement of spin-allowed and spin-forbidden states is opposite in the bilayer for an indirect exciton transition at 1.5 eV, where the spin-forbidden dark exciton is the lowest and the splitting energy is ~ 3 meV. This excludes, as a significant factor, the temperature change in the lattice parameter of the substrate,[62] which should provide similar effect in a monolayer and bilayer. Note that such arrangement with the lowest spin-forbidden state in the indirect transition series was previously proposed for multilayer $MoS_2$ structures based on the temperature dependencies of the 1.3-eV radiation.[24,80]

The proposed band structure variation under strain is in good agreement with the TRPL data. In particular, the absence of the slowly decaying PL component (with time $t_3$) in the radiation at the A exciton energy in unstrained monolayers indicates that the momentum indirect transition is impossible here. Its appearance under compression reflects a transformation of the band structure when the Λ-K exciton turns out to be very close to the K-K one (compare Figure 8 (a) and (c)). Analogous to the splitting $\Delta_{AF}$ of the spin- allowed and forbidden exciton states, this crossover can be characterized as a direct-indirect energy splitting $\Delta_{DI}$ ($E_0$-$E_c$ in Figure 3 (a)), which is about 4 meV in our samples. In the highly strained bilayer, which already has the momentum-indirect K−Γ transitions at 1.5 eV, the similar transformation is hardly possible for the scheme depicted in Figure 9 (d). It explains the different dependencies of the contributions to the integral intensities shown in Figures 6 (d) and 8 (d).

We emphasize that some contribution of the trion emission, if any exists, cannot explain the increasing dependences of the characteristic decay times measured in our samples (see Figure 6). The characteristic decay time of the trions in a TMD monolayer is close to that of a neutral exciton.[84] When the trion decay time is a bit longer,[74] it rapidly decreases to the exciton value with increasing temperature due to the rapid dissociation of the trion.

Fine structure of the trion states in 2D systems was analyzed in several papers.[85, 86, 87] In general, the trions heritage the symmetry properties of the neutral excitons. In the paper by Arora et al.,[42] it was found that the dark trion is 2 ±3 meV above the bright trion in $MoS_2$/hBN on sapphire substrate that is consistent rather with our results, than with the data of magneto-spectroscopy reported by Robert et al.[43] It should be noted that we consider the characteristics of the exciton states driven by spin or momentum separately. Such an approximation allows us to outline the variation range of parameters required for a more sophisticated model. Some important data derived from the performed experiments are given in **Table 1**.

**Table 1.** Values of strain, PL energies at low (LT) or room (RT) temperatures, bright-dark exciton splitting, $\Delta_{AF}$, derived for the A exciton and indirect IX exciton in $MoS_2$ MLs and BLs.

| Type of $MoS_2$ nanostructure | Strain $\varepsilon_{//}$ [%][a] | $\Delta_{AF}$ for A exciton [meV][c] | $\Delta_{AF}$ for IX [meV][c] |
|---|---|---|---|
| ML on $SiO_2$/Si | 0 | -1.5 [c] | |
| ML on $Al_2O_3$ | -0.12 | -4 | |
| BL on $Si_3N_4$/Si | 0 | 0 | +2 |
| BL on $Al_2O_3$ | -0.28 | -10 | +3 |
| ML/hBN | +0.06 [b] | +14 [c] | |
| BL/hBN | +0.29 [b] | | |

a) The $\Delta_{AF}$ values in this work are determined with an accuracy not higher than ±1 meV; b) strain values by Han et al;[20] c) $\Delta_{AF}$ value by Robert et al. [43]

In addition, we present in Table 1 the latest literature data on the bright-dark exciton splitting energy in the hBN-encapsulated $MoS_2$ monolayer obtained by magneto-PL spectroscopy.[43] According to the results presented in Ref.[20], the encapsulated monolayer should undergo tensile strain, whose absolute magnitude $|\varepsilon_{//}|$ is almost the same as in our structures under compression. It is noteworthy that in the $MoS_2$/hBN the magnitude of $\Delta_{AF}$~14 meV is of the same order as in our samples, but the sign of $\Delta_{AF}$ is opposite, i.e. the dark exciton is the lowest. Assuming a linear dependence of $\Delta_{AF}$ on the $|\varepsilon_{//}|$ value, we can conclude that deformation of the opposite sign can give a similar effect in terms of the magnitude of its impact, but will shift the exciton levels in opposite directions. This phenomenon arises likely due to the complex nature of the spin-orbit coupling, the two components of which, associated with chalcogen and metal orbitals, almost cancel each other in unstrained 2D $MoS_2$,[88] while strain can disturb this delicate balance. We believe that possible dependence of the dark-bright exciton splitting on strain can explain the huge dispersion of the data on the fine exciton spectrum in 2D $MoS_2$, published in literature.[4,41–43]

It can also be noted that the characteristic lifetimes for dark excitons are not as long as in quantum dots or in epitaxial 2D GaN/AlN systems,[48,81] where they can approach tens and hundreds of ns at low temperatures. This fact, which is probably associated with the different energy splitting between the dark and bright excitons, along with the strong oscillator strength in the TMDs, requires a comprehensive study along with the dependence of the fine spectrum of the exciton states on strain.

## 6. Conclusion

We carried out time-resolved micro-PL measurements in atomically thin $MoS_2$ films transferred to different substrates, which creates different strains inside them. The main results of this work are the elucidation of the fine exciton spectrum and the discovery of its dependence on strain. The change of fast, intermediate, and slow decaying components with temperature, their characteristic times and contributions to the total radiation, makes it possible to estimate the spin allowed-forbidden splitting $\Delta_{AF}$ and to assume the existence of transitions indirect in momenta near the A exciton energy due to the strain-induced modification of the band structure. Our findings clearly show that the bright exciton in the A exciton series (1.9 eV) is the lowest in both the unstrained monolayer and bilayer, and the splitting energy does not exceed 2 meV. The splitting is almost equal to zero in the unstrained bilayer due to constraints associated with the symmetry of an even number of monolayers. It is also shown that the bright-dark exciton splitting value increases several times with increasing compressive strain. On the contrary, the dark exciton is the lowest in the indirect exciton series (1.5 eV) in bilayers. Such an arrangement in the bilayer leads to an additional decrease in the light emission intensity for this indirect transition in the system with two competing channels of recombination. We have determined the decay time constants characteristic of the emission of spin- and momentum-forbidden states; at low temperatures they are in the range (0.2–0.4) ns and (6–7) ns, respectively. The observed trends in variation of the fine exciton spectrum upon strain allow us to suggest that the sign of splitting could be opposite with tensile deformation. We underline that we have used a simplified approach, taking into account only the basic channels of recombination. To refine the results obtained, it is necessary to consider other factors, for example, nonradiative recombination and the contribution of excitons which occur out of the light cone. We hope that our results will stimulate studies using angular-resolved PL spectroscopy and magneto-PL of structures under different strain. This work demonstrates that the effect of strain must be taken into account in order to obtain the desired optical properties in atomically thin TMD nanostructures.

## 7. Methods

*Sample preparation*

Atomically thin $MoS_2$ flakes were mechanically exfoliated from the bulk crystal (production of HQ Graphene) and transferred by dry viscoelastic stamping[89] using a commercial HQ Graphene 2D transfer system at the Ioffe Institute and homemade one at the TU Ilmenau. To form planar structures, the flakes were positioned on a $SiO_2/Si$ substrate or a $Si_3N_4/Si$ one. The surface roughness of both $MoS_2/SiO_2$ and substrate was around 0.3 nm. We emphasize that this level of substrate roughness was optimal for our study. Atomically-thin layers lay freely, without tight contact with the substrate. As a result, the strains in our planar samples were close to zero. Sapphire substrates with pyramids on their surface were used to create stress in the flakes suspended from the tops of the pyramids. To realize the adhesion between the flakes and the substrate, the $MoS_2$ flakes attached to the Gel-Pak PF X4 polymer film came into contact with the substrate while the glass holder was tilted at a small angle of 1-2° with respect to the substrate. When full contact was achieved between the flake and the substrate, the holder was moved a few microns in the xy-direction and then slowly lifted in the z-direction. This procedure made it possible to obtain a uniform film of a suitable size – more than $10 \mu m^2$ for monolayer and bilayer regions. Based on the dependencies reported by Amani et al.[90] we conclude from the optical characterization that internal quantum yield in our structures is about 1%.

*Micro-Raman and cw micro-PL measurements*

Micro-Raman measurements were carried out using a Horiba Jobin Yvon T64000 spectrometer with 1800 gr/mm diffraction grating. At room temperature, the laser beam ($\lambda = 532$ nm) incident normally to the surface was focused by Olympus MPLN 100x (NA = 0.9) objective into a spot with a diameter of less than 1 μm. In order to suppress the Rayleigh scattering and to obtain information from the low-frequency (5-50 $cm^{-1}$) range, a set of BragGrate optical filters was used. Micro-PL measurements at cw excitation were carried out on the same spectrometer using a 600 gr/mm grating. For low-temperature measurements, a Linkam THMS600 temperature-controlled microscope stage and a long-working-distance Leica PL FLUOTAR 50x (NA = 0.55) objective was used. To avoid laser-induced modification of the $MoS_2$ films, the laser power was limited to 400 μW.

*Micro-PL measurements with temporal resolution (micro-TRPL)*

The micro-PL spectra and PL decay curves of $MoS_2$ samples were measured at temperatures ranging from 8-10 K to 300 K using the ST-500-Attocube cryostat (Janis) supplied with a temperature controller. The sample was adjusted with an accuracy of about ~20 nanometers, using a three-coordinate piezo-driver located directly in the cold zone of the cryostat. This provides mechanical stability and vibration isolation during prolonged temperature–dependent measurements. PL excitation was carried out by focusing laser radiation on a sample with a minimum spot size of the order of 3-5 μm. Focusing was achieved by introducing laser radiation into an objective (Mitutoyo plan apochromat with 50x magnification, NA = 0.42 and focal length 4 mm), which was also used to collect PL radiation. The PL radiation transmitted through the objective was focused by a triplet achromatic lens in the plane of the mirror with a calibrated aperture (Pinhole) with a diameter of 200 μm. Due to the magnification of the objective, this aperture corresponds to a resolution of ~4 μm. The photoluminescence radiation passed through the aperture was collected and focused onto the entrance slit of an SP-2500 spectrometer (Princeton Instruments) with a grating 600 gr/mm using two triplet achromatic lenses. For additional blocking of laser radiation scattered on the sample surface and/or optical elements, a bandpass interference filter was used. A cooled PyLoN CCD (Princeton Instruments) was used as a PL detector in the spectrometer. To measure the micro-TRPL spectra in this work, we used a picosecond pulsed semiconductor laser PILAS 405 nm (Advanced Laser Systems) with the average excitation power of 100 nW measured before the cryostat window. A single-photon avalanche photodiode (SPAD) PDM 100 (Micro Photon Devices) with time resolution ~40-50 ps was chosen as a detector for TR measuring. The time-correlated single photon counting system SPC-130 (Becker & Hickl) was used. To isolate a chosen excitonic line from emission of the background and other excitonic lines we used long-pass and short-pass tunable interference optical filters.

**Supporting Information**

Supporting Information is available from the corresponding author upon a reasonable request.

Acknowledgements

This work was supported by the Russian Science Foundation (project #19-12-00273). Samples fabrication was partly funded by DAAD #57435564 Carl Zeiss Foundation P2018-01-002 and TAB 2018 FGR 0088. The authors are grateful to M. Glazov, A. Rodina, and M. Nestoklon for fruitful discussions, B. Borodin and L. Kotova for the help in characterization of samples.

I. A. Eliseyev, A. I. Galimov, M. V. Rakhlin, E. A. Evropeitsev, A.°A.°Toropov, V. Yu. Davydov, S. Thiele, J. Pezoldt, and T.°V.°Shubina*

**Photoluminescence kinetics of dark and bright excitons in atomically thin MoS₂**

Time-resolved micro-photoluminescence and micro-Raman studies of 2D MoS2 show that strain can dramatically affect the fine spectrum of exciton states. In unstrained monolayers, the spin-allowed A exciton is the lowest with the dark-bright exciton splitting <2 meV, while compressive strain increases this value by several times. In contrast, the spin-forbidden state is the lowest in the indirect exciton series in bilayers.

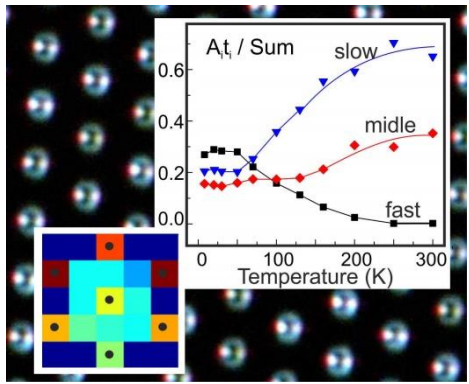